\documentstyle[aps,manuscript]{revtex}
\tightenlines
\draft
\begin{document}
\title{Cosmological expansion governed by a scalar field
from a 5D vacuum}
\author{
$^{1,2}$Mauricio Bellini\footnote{
E-mail address: mbellini@mdp.edu.ar}}
\address{
$^1$Departamento de F\'{\i}sica,
Facultad de Ciencias Exactas y Naturales,
Universidad Nacional de Mar del Plata,
Funes 3350, (7600) Mar del Plata, Argentina.\\
$^2$ Consejo Nacional de Ciencia y Tecnolog\'{\i}a (CONICET).}

\vskip .5cm
\maketitle
\begin{abstract}
We consider a single field governed
expansion of the universe from a five
dimensional (5D) vacuum
state. Under an appropiate change of variables the universe
can be viewed in a effective manner as expanding in 4D with
an effective equation of state which describes different epochs
of its evolution. In the example here worked the universe fistly
describes an inflationary phase, followed by a decelerated
expansion. Thereafter, the universe is accelerated and describes
a quintessential expansion to finally, in the future, be
vacuum dominated.
\end{abstract}
\vskip .2cm
\noindent
Pacs numbers: 04.20.Jb, 11.10.kk, 98.80.Cq \\
\vskip 1cm
\section{Introduction and Overview of the 5D formalism}

In the last years has been an uprising interest in finding exact solutions
of the Kaluza-Klein field equations in 5D, where the
fifth coordinate is considered as noncompact\cite{1}.
Unlike the usual Kaluza-Klein theory in which a cyclic symmetry associated
with the extra dimension is assumed, the new approach removes the cyclic
condition on the extra dimension and derivatives of the metric with
respect to the extra coordinate are retained. This induces non-trivial
matter on the hypersurfaces with $\psi = constant$ and other
nontrivial frames.
This theory reproduces
and extends known solutions of the Einstein field equations in 4D. Particular
interest revolves around solutions which are not only
Ricci flat\cite{2a}, but also Riemann flat\cite{2b}: $R^A_{BCD}=0$ (
$A$, $B$, $C$, $D$ can take the values $0,1,2,3,4$).
This is because it is
possible to have a flat 5D manifold which contains a curved
4D submanifold, as implied by the Campbell theorem\cite{3}. So, the
universe may be ``empty'' and simple in 5D, but contain matter of
complicated forms in 4D\cite{we}.

One of the greatest challenges of modern cosmology is understanding the
nature of the observed late-time acceleration of the universe. Recent
measurements of type Ia
Supernovae (SNIa)\cite{...} at redshifts $z \sim 1$ and also the
observational results coming from the Cosmic Microwave Background
Radiation (CMBR) along with the Maxima\cite{..} and Boomerang data\cite{.}
indicate that the expansion of the present universe is accelerated.
In fact the present day results show that supernovae are moving
faster than
expected from the luminosity redshift relationship in a decelerating
universe. A possible explanation is that in the universe there exists
an important matter component which, in its most simple description,
has the characteristic of a cosmological constant as vacuum energy
density which contributes to a large component of negative
pressure\cite{bertolami,bellini}. The idea that the expansion
of the universe could be governed by a scalar field
has been developed in the quintessential
models, where the dynamics of the scalar field is governed by 
an appropiate potential $V(\varphi)$\cite{stein}.
This paper is devoted to study the evolution of the universe
which firstly suffers an inflationary expansion followed
by a decelerated expansion (matter and radiation dominated) to finally
be monotonically accelerated until a quasi-de Sitter expansion from
a 5D vacuum state, where the (space-like) fifth dimension is considered
as noncompact.
We shall suppose that the universe is governed by a scalar
field, which is minimally coupled to gravity. For the system,
we shall consider that the action is
\begin{equation} \label{action}
I=-\int d^{4}x \  d\psi\,\sqrt{\left|\frac{^{(5)} g}{^{(5)}g_0}
\right|} \ \left[
\frac{^{(5)} R}{16\pi G}+ ^{(5)}{\cal 
L}(\varphi,\varphi_{,A})\right],
\end{equation}
where $\varphi$ is a scalar (neutral) quantum field,
$G=M^{-2}_p$ is the gravitational constant
[being $M_p=1.2 \  10^{19} \  {\rm GeV}$ the Planckian mass],
and $^{(5)}R$ is the Ricci scalar.
Furtheremore $^{(5)} g$ is the determinant of the covariant tensor
metric $g_{AB}$ ($A,B$ can take the values $0,1,2,3,4$).
We are interested to describe a manifold in apparent vacuum,
so that the Lagrangian density ${\cal L}$ in (\ref{action}) should be only
kinetic 
\begin{equation}\label{lagr}
^{(5)} {\cal L}(\varphi,\varphi_{,A}) = \frac{1}{2} g^{AB}
\varphi_{,A} \varphi_{,B}.
\end{equation}
To describe a 3D spatially isotropic and homogeneous universe, which
is Ricci flat: $R^A_{BCD}=0$ and describes a 5D vacuum: $G_{AB}=0$,
we shall consider the background metric\cite{plb2004}
\begin{equation}\label{a}
dS^2 = \psi^2  dN^2 - \psi^2 e^{2N} dr^2 - d\psi^2,
\end{equation}
where $\psi$ is the space-like fifth coordinate, $r(x,y,z)
=\sqrt{x^2+y^2+z^2}$ and
$N$, $x$, $y$, $z$ are dimensionless
coordinates.
Furthermore, the metric (\ref{a}) is 3D spatially isotropic, homogeneous
and flat.
For the metric (\ref{a}), the determinant of the covariant
metric tensor $g_{AB}$ is $|^{(5)}g|=\psi^8 e^{6N} $
and $|^{(5)} g_0|= \psi^8_0 e^{6N_0}$
is a constant of dimensionalization, for the constants $\psi=\psi_0$
and $N=N_0$.
On the other hand, the energy-momentum tensor is given by
\begin{equation}
T_{AB} = \varphi_{,A} \varphi_{,B}-
\frac{1}{2} g_{AB} \varphi_{,C} \varphi^{,C},
\end{equation}
which is symmetric because the symmetry of $g_{AB}$.
The dynamics for $\varphi$ is described
by the Lagrange equation
\begin{equation}\label{lag}
\frac{\partial^2\varphi}{\partial N^2}
+ 3\frac{\partial\varphi}{\partial N} - e^{-2N} \nabla^2_r \varphi -
\psi\left(\psi \frac{\partial^2\varphi}{\partial\psi^2}
+ 4 \frac{\partial\varphi}{\partial \psi}\right)=0.
\end{equation}
In absence of 3D spatially isotropic field fluctuations, the background
field $\varphi_b(N,\psi)$ corresponding to the vacuum equation
$\left({\partial\varphi_b\over \partial N}\right)^2 + \psi^2
\left({\partial\varphi_b\over\partial\psi}\right)^2 = \left.T_{AB}\right|_b
=0$ on the background metric (\ref{a}), is a constant of $N$ and $\psi$:
$\varphi_b(N,\psi)=const.$.
The commutator between $\varphi$ and $\stackrel{\star}{\varphi}$ will
be
\begin{equation}\label{commut}
\left[\varphi(N,\vec r,\psi), \stackrel{\star}{\varphi}
(N,\vec{r'},\psi')\right] =
i \left|\frac{^{(5)} g_0}{^{(5)} g}\right| \  \delta^{(3)}(\vec r-
\vec{r'}) \delta(\psi - \psi').
\end{equation}
In order to simplify the structure of the equation (\ref{lag})
we can make the transformation $\varphi =\chi
e^{-3N/2}\left(\psi_{0}\over \psi \right)^{2}$.
The dynamics fo $\chi (N,\vec{r},\psi)$ being described
by the equation of motion
\begin{equation}\label{mot}
\stackrel{\star \star}{\chi}-\left[e^{-2 N}\nabla_{r}^{2} 
+\left(\psi^{2} \frac{\partial^2}{\partial\psi^2} +\frac{1}{4}\right) 
\right]\chi =0, 
\end{equation}
and
the commutator between $\chi$ and
$\stackrel{\star}{\chi}$ is
\begin{equation}\label{comm}
\left[\chi(N,\vec r,\psi), \stackrel{\star}{\chi}
(N,\vec{r'},\psi')\right] =
i  \delta^{(3)}(\vec r- \vec{r'}) \delta(\psi - \psi'). 
\end{equation}
We can make a Fourier expansion for $\chi$
\begin{eqnarray} \label{ec4}
\chi (N,\vec{r},\psi)&=& \frac{1}{(2\pi)^{3/2}}\int d^{3} k_{r} \int d 
k_{\psi} \left[a_{k_{r}k_{\psi}} e^{i(\vec{k_r} \cdot \vec{r} 
+{k_\psi}\cdot {\psi})}\xi_{k_{r}k_{\psi}}(N,\psi) \right. \nonumber
\\
&+& \left. a_{k_{r}k_{\psi}}^{\dagger} e^{-i(\vec{k_r} \cdot \vec{r} 
+{k_\psi} \cdot {\psi})}\xi_{k_{r}k_{\psi}}^{*}(N,\psi)\right],
\end{eqnarray}
where the asterisk denotes the complex conjugate and 
$(a_{k_{r}k_{\psi}},a_{k_{r}k_{\psi}}^{\dagger})$ are respectively
the annihilation
and creation operators. They satisfy the following commutation
expressions
\begin{eqnarray}\label{ec5}
\left[a_{k_{r}k_{\psi}},a_{k_{r}k_{\psi}}^{\dagger}\right]&=&\delta^{(3)
}\left(\vec{k_r}-\vec{k'_r}\right) 
\delta\left(\vec{k_\psi}-\vec{k'_\psi}\right),\\
\label{ec6}
\left[a_{k_{r}k_{\psi}}^{\dagger},a_{k'_{r}k'_{\psi}}^{\dagger}\right]
&=&\left[a_{k_{r}k_{\psi}},a_{k'_{r}k'_{\psi}}\right]=0.
\end{eqnarray}
The expression (\ref{comm}) complies if the modes are normalized
by the following condition:
\begin{equation} \label{recon}
\xi_{k_{r}k_{\psi}}\left(\stackrel{\star}{\xi}_{k_{r} 
k_{\psi}}\right)^{*} - \left(\xi_{k_{r} k_{\psi}}\right)^{*} 
\stackrel{\star}{\xi}_{k_{r}k_{ \psi }} = i.
\end{equation}
This equation provides the normalization for the complete set of
solutions on all the spectrum ($k_r,k_{\psi}$).
As was demonstrated in a previous work\cite{plb2006},
$\xi_{k_r k_{\psi}}(N,\psi)=e^{-i \vec{k}_{\psi}.\vec{\psi}}
\bar\xi_{k_r}(N)$, where $\bar\xi_{k_r}(N)$ is a solution of
\begin{equation}\label{mod5D}
\stackrel{\star\star}{\bar\xi}_{k_r} + \left( k^2_r e^{-2N}
-\frac{1}{4}\right)
\bar\xi_{k_r}=0,
\end{equation}
such that the normalization condition for $\bar\xi_{k_r}(N)$ becomes
\begin{equation}
\bar\xi_{k_{r}}\left(\stackrel{\star}{\bar\xi}_{k_{r} 
}\right)^{*} - \left(\bar\xi_{k_{r}}\right)^{*} 
\stackrel{\star}{\bar\xi}_{k_{r}} = i,
\end{equation}
where the overstar denotes the derivative with respect to $N$.
Hence, the field $\chi$ in eq. (\ref{ec4}) can be
rewritten as
\begin{equation}\label{chichi}
\chi(N,\vec r)
=\frac{1}{(2\pi)^{3/2}} {\Large\int} d^3k_r
{\Large\int} dk_{\psi} \left[ a_{k_r k_{\psi}} e^{i \vec{k_r}.\vec r}
\bar\xi_{k_r}(N) + a^{\dagger}_{k_r k_{\psi}} e^{-i \vec{k_r}.\vec r}
\bar\xi^*_{k_r}(N)\right].
\end{equation}
Furthermore, the functions $\bar{\xi}_{k_r}(N)$
are given by\cite{plb2006}
\begin{equation}
\bar{\xi}_{k_r}(N)
=\frac{i \sqrt{\pi}}{2} {\cal H}^{(2)}_{1/2}
\left[ k_r e^{-N}\right].
\end{equation}
Finally, the field $\varphi$ is given by
\begin{equation}
\varphi(N,\vec r,\psi) = e^{-\frac{3N}{2}}
\left(\frac{\psi_0}{\psi}\right)^2
\chi(N,\vec r),
\end{equation}
with $\chi(N,\vec r)$ given by eq. (\ref{chichi}).
It is very important to notate
that exponentials $e^{\pm i \vec{k}_{\psi}.\vec{\psi}}$ disappear
in $\chi(N,\vec r)$ and there is not dependence on the fifth coordinate
$\psi$ in this field. This fact is an evidence
of that the field $\varphi(N, \vec r, \psi)$ propagates only on the
3D spatially isotropic space $r(x,y,z)$, but not on the additional
space-like coordinate $\psi$.

\section{An effective 4D model of expansion for the universe}

In order to develope an effective 4D model for an universe which
is governed by the scalar field $\varphi$, we shall use the following
change of variables on the metric (\ref{a})
\begin{equation}\label{trans}
dt= \psi \  dN, \qquad dR=\psi \  dr, \qquad \psi = \psi
\end{equation}
so that the resulting metric becomes
\begin{equation}              \label{me}
dS^2 = dt^2 - e^{2\int \frac{dt}{\psi}} dR^2 - d\psi^2.
\end{equation}
On hypersurfaces $\psi=1/H(t)$ the metric (\ref{me}) becomes
\begin{equation}\label{a1}
dS^2=dt^2 - e^{2 \int H dt} dR^2 - [d\left(H^{-1}\right)]^2,
\end{equation}
which also can be rewritten as an effective 4D metric
\begin{equation}\label{m4d}
dS^2 \rightarrow ds^2= \left[1-\left(\frac{\dot H}{H^2}\right)^2\right]
dt^2 -  e^{2 \int H dt} dR^2,
\end{equation}
which is well defined when $1-\left({\dot H\over H^2}\right)^2 \neq 0$.
Notice that, from the mathematical point of view,
the change of variables (\ref{trans}) do not describes
transformation of coordinates. This change of variables
describes a map from the particular frame $U^{\psi}=0$ [of the
metric (\ref{a})], to the particular frame $u^R=0$ of the effective
4D metric (\ref{m4d}). Here, $U^A={dx^A \over dS}$
are the penta velocities of the original metric (\ref{a}) and
$u^{\mu} = {dX^{\mu} \over ds}$ are the tetra velocities of the
effective 4D metric (\ref{m4d}), such that $g^{AB} U^A U^B=1$
and $g^{\mu\nu} u^{\mu} u^{\nu} =1$, respectively.

Furthermore, the density Lagrangian ${\cal L}$ of eq. (\ref{lagr}) can
be expanded as
\begin{equation}
{\cal L}(\varphi,\varphi_{,A}) = \frac{1}{2} \left[
g^{\mu\nu} \varphi_{,\mu}\varphi_{,\nu} + g^{\psi\psi} \varphi_{,\psi}
\varphi_{,\psi}\right],
\end{equation}
so that we can make the following identification for the
scalar potential on the effective 4D metric (\ref{m4d}):
\begin{equation}\label{pot}
V(\varphi) = \left.\frac{1}{2} g^{\psi\psi} \varphi_{,\psi}
\varphi_{,\psi}\right|_{\psi=1/H} = 2 H^2 \varphi^2(t,\vec R).
\end{equation}
Using the change of variables (\ref{trans}),
the equation (\ref{lag}) can be
rewritten on the 4D effective submanifold (\ref{m4d}), as
\begin{equation}\label{23}
\ddot\varphi + \left(3H - \frac{\dot H}{H}\right) \dot\varphi
- e^{2 \int H dt} \nabla^2_r\varphi + V'(\varphi)=0,
\end{equation}
with
\begin{equation}
V'(\varphi) = - H^2 \left.\left[ \psi^2 \varphi_{,\psi\psi} + 4
\psi \varphi_{,\psi}\right]\right|_{\psi=H^{-1}}= 2 H^2\varphi.
\end{equation}

\subsection{The effective 4D equation of state}

Using the effective 4D metric (\ref{m4d}), we obtain the
following Einstein's equations
\begin{eqnarray}
&& \frac{3 H^6}{\dot H^2 - H^4} = 8\pi G \rho, \\
&& \frac{H^6}{\left(\dot H^2 - H^4\right)^2} \left[
3H^4 - 3\dot H^2 + 2\dot H H^2 -6\frac{\dot H^3}{H^2} +
2 \ddot H \frac{\dot H}{H}\right] = 8\pi G {\rm p},
\end{eqnarray}
where $\rho$ and ${\rm p}$ are, respectively, the energy density and the
pressure.
The effective 4D equation of state for the universe
becomes
\begin{equation}\label{state}
\frac{{\rm p}}{\rho} = \omega_{eff}(t)
\end{equation}
with
\begin{equation}\label{omegae}
\omega_{eff}(t) = -\left[1+ \frac{\left(2\dot H H^2 - 6\frac{\dot H^3}{H^2} +
2\frac{\dot H \ddot H}{H}\right)}{3\left(H^4 - \dot H^2\right)}\right],
\end{equation}
which, for a Hubble parameter $H=p/t$ (with constant $p$) agrees
exactly with that of a spatially flat
4D Friedmann-Robertson-Walker (FRW) metric $ds^2 = dt^2 - e^{2\int H dt}
dR^2$:
$\left.\omega_{eff}(t)\right|_{H=p/t} = {2-3p\over 3p}=\omega_{FRW}(t)$.
Some particular cases $p\rightarrow \infty$, $p=2/3$ and $p=1/2$,
give us respectively $\omega_{eff} \rightarrow -1$, $\omega_{eff} =0$
and $\omega_{eff}=1/3$, which describes respectively
expansions dominated
by vacuum, matter and radiation.
Note that for $p >1$ the effective 4D metric (\ref{m4d}) is
always Lorentzian in nature. However for $p=1$ this metric is not
well defined. On the other hand, for $p(t) <1$ the metric (\ref{m4d})
is Euclidean and hence losses its relativistic nature.
Hence, a well defined model for the expansion of the universe must
be developed using $p > 1$ in the metric (\ref{m4d}).
For a more realistic model with a time dependent $p(t)$,
one obtains
$\left.\omega_{eff}(t)\right|_{H=p(t)/t}\neq \omega_{FRW}(t)$.
In the section (\ref{evol}) we shall study with more detail
this last case.

\subsection{Comoving frame}

In order to describe the evolution of the universe on a
comoving frame, which is the relevant for cosmological models,
we can make use of the hyperbolic condition $g_{\mu\nu} u^{\mu} u^{\nu}=1$
on the 4D effective metric (\ref{m4d}). Here, $u^{\mu} = {dx^{\mu}\over
dS(t)}$ ($\mu$ can take the values $0,1,2,3$),
are the tetravelocities being $u^R =0$ in the comoving frame.
In this frame the velocity $u^t$ is
\begin{equation}\label{ut}
u^t = \frac{1}{\sqrt{1-(\dot H/H^2)^2}}.
\end{equation}
Note that when
\begin{equation}\label{condicion}
-\dot H/H^2 \ll 1,
\end{equation}
we obtain $u^t \simeq 1$ and
the metric (\ref{m4d}) describes an asymptotic 4D FRW metric, which
usually is used to describe the universe in cosmological models.

\subsection{Evolution of the background field $\varphi_b$}

The effective 4D evolution of the background field $\varphi_b(t)$ is given
by the equation (\ref{23}) with $\nabla^2_R\varphi=0$
\begin{equation}\label{varphib}
\ddot\varphi_b + \left(3H - \frac{\dot H}{H}\right)\dot\varphi_b +
2 H^2 \varphi_b =0,
\end{equation}
which has the general solution
\begin{equation}\label{gsol}
\varphi_b(t) = \frac{\varphi_b(0)}{2} e^{-\int H(t) dt}
\left[1+ e^{-\int H(t) dt}\right],
\end{equation}
where $\varphi_b(0) = \varphi_b(t=0)$. It is evident that $\varphi_b(t)$
decreases monotonically with the time and rolls down the minimum of the
potential, so that $\varphi_b(t\rightarrow \infty) \rightarrow 0$.

\section{Model of expansion of the universe}\label{evol}

In order to describe all the evolution of the universe
we can propose the following expresion for the Hubble: $H(t)=p(t)/t$, such
that
\begin{equation}\label{pw}
p(t) = 1.8 a t^{-n}
+ \left(\frac{b^2}{4a} +0.62\right)
+ c t
\end{equation}
where
$a = 1/6 \  10^{30n} \  G^{n/2}$, $b=8/7 \  10^{15n} \  G^{n/4}$,
$c=2.0 \  10^{-61} \  G^{-1/2}$ and $n=0.352$.
The general solution (\ref{gsol}) in the example we are worked assume
the explicit expression
\begin{equation}\label{vb}
\varphi_b(t) = \varphi_b(0) e^{-\frac{3 c t}{2}} \left(\frac{t}{t_0}
\right)^{-31/25} \left\{ \left(\frac{t}{t_0}\right)^{-b^2/(2a)}
e^{-\left[\frac{36 a t^{-n} - 5 c n t}{10 n}\right]} +
\left(\frac{t}{t_0}\right)^{\frac{62 a-25 b^2}{100 a}}
e^{\frac{18 a t^{-n} + 5 c n t}{10 n}}\right\}.
\end{equation}
In the figure (1) is ploted the parameter $p[x(t)]$ [where
$x(t)={\rm log}_{10}(t)$], which
always remains with values $p>1$. In the figure (2) is shown
the evolution for the cosmological parameter $w_{eff}[x(t)]$. Note
that for $x(t)<20$ the universe is governed by vacuum and describes
an inflationary expansion, but later
$w_{eff}[x(t)]$ increases to thereafter describe a phase with positive
pressure: $\omega_{eff} > 0$
(in the range $ 30 < x(t) < 60$). After it, the pressure decreases and
for $x(t) > 60$ the universe expands with
negative pressure
until the present day, when $\omega_{eff}[x\simeq 60.652] \simeq -0.7$.
This result agrees with the experimental data\cite{ex}.
In the figure (3) is ploted $(\omega_{eff}-\omega_{FRW})[x(t)]$.
It is very clear that the discrepance between $\omega_{eff}$ and
$\omega_{FRW}$ becomes more notorious in epochs where
the universe expands with positive pressure.
However, in both epochs (in the very early universe
and the present day universe), the equation of state for the universe
agrees with that predicted by a 4D spatially flat FRW metric: $ds^2=
dt^2 - e^{2\int H dt} dR^2$. More exactly, the condition (\ref{condicion})
holds for
\begin{equation}\label{condicion1}
\frac{1}{p(t)} - \frac{t \dot p(t)}{p(t)^2} \ll 1.
\end{equation}

\section{Final Comments}

In this letter we have studied the evolution of the universe which
is considered as governed by a single scalar field from a 5D vacuum
state, where the fifth dimension is considered as noncompact [see
the metric (\ref{a})]. However when we make the change of variables
(\ref{trans}) the universe describes an effective 4D evolution
which can be described by the metric (\ref{m4d}).
This metric is well defined for $\left(\dot H/H^2\right)^2 \neq 1$,
which is the case we are considered in this letter.
If we consider
a Hubble paramter $H(t) = p(t)/t$, this metric remains Lorentzian
for $p >1$. In the example here considered, with a power law
expansion (\ref{pw}), the universe initially describes an
equation of state with $\omega_{eff} \simeq -1$, which increases
until take values of the order of $\omega_{eff} \simeq 0.4$. Thereafter,
this cosmological parameter begins to descrease to take values
which agree very good with the present day experimental data
$\omega_{eff} \simeq -0.7$. This result is consistent with
a quintessential expansion. The model predicts that, in the future,
the universe
will be more and more accelerated to finally describe
an effective 4D vacuum dominated expansion $\omega_{eff} \simeq -1$.

\vskip .3cm
\noindent
\centerline{{\bf Acknowledgements}}
\vskip .1cm
\noindent
MB acknowledges CONICET (Argentina) and UNMdP for financial
support.\\

\newpage
\begin{displaymath}
\end{displaymath}
\vskip 7cm
\noindent
{\rm Fig. 1)} Evolution of $p[x(t)]$
as a function of $x(t) = {\rm log}_{10}(t)$.\\
\vskip 7cm
\noindent
{\rm Fig.2)} Evolution of $\omega_{eff}$
as a function of $x(t) = {\rm log}_{10}(t)$.\\
\vskip 7cm
\noindent
{\rm Fig. 3)} Evolution of $\omega_{eff}-\omega_{FRW}$
as a function of $x(t) = {\rm log}_{10}(t)$.\\

\end{document}